\documentclass[prl,twocolumn,a4paper,aps,showpacs,superscriptaddress]{revtex4}
\usepackage{graphicx}
\usepackage{dcolumn}
\usepackage{bm}
\usepackage{amssymb}
\usepackage{epsfig}
\begin{document}
\title{Superconducting double spin valve with  extraordinary large tunable magnetoresistance}
\author{Francesco Giazotto}
\email{f.giazotto@sns.it}
\affiliation{NEST CNR-INFM and Scuola Normale Superiore, Piazza dei
Cavalieri 7, I-56126 Pisa, Italy}

\begin{abstract}
A superconducting double spin valve device is proposed. Its operation takes advantage of the interplay between the spin-filtering effect of \emph{ferromagnetic insulators} and superconductivity-induced out-of-equilibrium transport. Depending on the degree of nonequilibrium, extraordinary large tunnel magnetoresistance as large as $10^2...10^6\%$ can be obtained for realistic material parameters, and it can be tuned over several orders of magnitude under proper voltage biasing and temperature. The relevance of this setup for low-temperature applications is further discussed.
\end{abstract}
\maketitle

The possibility to generate highly spin-polarized electric currents is one the main topics in the field of \emph{spintronics} \cite{Zutic}, both from the fundamental \cite{Jedema,Giazotto} and the technological point of view \cite{Fert}. In this context, a widespread route to achieve spin-polarized transport is through tunneling magnetoresistance (TMR) \cite{Zutic} in magnetic tunnel junctions, which consist typically of ferromagnetic electrodes separated by a nonmagnetic insulating layer.
Magnetic tunnel junctions have recently demonstrated TMR values as large as $500\%$ at room temperature \cite{Lee}, and show nowdays an enormous technological impact for the realization of a wide range of devices spanning, for instance,  from magnetic random access memories (MRAM) to sensors.
On the other hand, tunneling through a \emph{ferromagnetic insulator} (FI) \cite{Meservey,Miao,Moodera2} from \emph{nonmagnetic} electrodes is an alternative and effective choice to produce highly spin-polarized currents. Here one takes advantage of the spin-filtering property of FIs due to the different barrier heights experienced by the two spin species \cite{Worledge,Filip,Xie,Saffarzadeh}.
Spin filtering efficiency approaching $100\%$ has been achieved in tunnel junctions comprising Eu-based FIs \cite{Moodera3,LeClair,Moodera4,Santos}, thus showing the potential of these materials for spintronic applications.

Here we propose a double spin valve device which combines the spin-filtering effect of FIs with superconductivity-induced nonequilibrium. Extraordinary large TMR ($\sim 10^2...10^6\%$) can be achieved, which can be tuned over several orders of magnitude under voltage biasing and temperature.
Such features make
this structure attractive for the implementation of magnetoresistive nanodevices in superconducting spintronics.

\begin{figure}[b!]
\includegraphics[width=7.5cm]{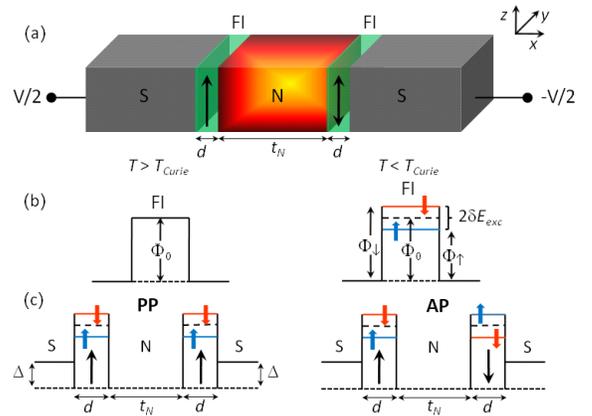}
\caption{\label{fig1} (Color) (a) Double-barrier tunnel structure consisting of two superconductors (S) and a normal metal (N) separated by thin ferromagnetic insulators (FI). (b)  FI  potential profile for $T>T_{Curie}$ (left) and $T<T_{Curie}$ (right)
(see text).
(c) Schematic energy-band diagram of the double-barrier structure in the parallel (PP) and antiparallel (AP) alignment of magnetizations in FIs. $\Delta$ is the superconducting energy gap.}
\end{figure}

The system we envision consists of two equal superconductors (S) symmetrically connected to a normal-metal layer of thickness $t_N$ through ferromagnetic insulating barriers (FI) [see Fig. 1(a)]. We assume the FI layers to be identical (a barrier asymmetry would not change the overall physical picture), $d$ labels the barriers width,
and a voltage $V$ is applied across the structure.
The magnetization of the left barrier is aligned and pinned along the $z$ axis, while the other can be freely switched by applying a small magnetic field \cite{interlayer} (coercitive fields as low as a few tens of Oe can be achieved, e.g., with EuS \cite{Kowalczyk}).
Moreover, we assume the N-layer resistance to be much smaller than the FI resistance, so to neglect any spatial variation of the spin accumulation within the N region, and that $t_N$ is smaller than the spin relaxation length in N.
Figure 1(b) displays the FI potential profile for temperatures ($T$) above (left) or below (right) the Curie temperature ($T_{Curie}$).
While in the former case electrons with opposite spin experience the same barrier height $\Phi_0$, for $T<T_{Curie}$ the FI conduction band splits leading to two barriers with height $\Phi_{\sigma}=\Phi_0\mp\delta E_{exc}$, where $\sigma=\uparrow(\downarrow)$ for up(down) spin, and $2\delta E_{exc}$ is the total splitting of the conduction band \cite{Meservey,Moodera3}.
The spin-dependent transmission probability through a FI barrier in WKB approximation is given by $\mathcal{T}_{\sigma}\cong e^{-2d\kappa_{\sigma}}$ \cite{Wolf}, where $\kappa_{\sigma}=\sqrt{2m\Phi_{\sigma}/\hbar^2}$ and $m$ is the free-electron mass.
Since each barrier resistance $R_{\sigma}$ is spin dependent ($R_{\sigma}\propto \mathcal{T}_{\sigma}^{-1}$), $R_{\downarrow}/R_{\uparrow}=e^{2d\delta \kappa}$ with $\delta \kappa=\kappa_{\downarrow}-\kappa_{\uparrow}$, so that spin-up current will largely exceed the spin-down one depending on $\delta E_{exc}$, leading to a high spin polarization.
Yet, in the present double spin valve the total current will  depend also on the relative direction of magnetizations in both FIs [see Fig. 1(c)] \cite{Worledge}. 
In particular, the total current flowing when the magnetizations are aligned in the parallel configuration (PP) will largely exceed that  flowing in the antiparallel alignment (AP).

In the sequential tunneling regime and within a simple relaxation-time approximation we can derive the steady-state spin-dependent  distribution functions in the N layer at finite bias $V$. In the PP as well as AP configurations they are given by
\begin{eqnarray}
f_{\uparrow}^{PP}(\epsilon,V,T)=\frac{\mathcal{N}_l f_l+\mathcal{N}_r f_r+f_0(e^{2d\delta\kappa}\Gamma_{\downarrow}\tau_{sf})^{-1}}{\mathcal{N}_l+\mathcal{N}_r+(e^{2d\delta\kappa}\Gamma_{\downarrow}\tau_{sf})^{-1}}\\
f_{\downarrow}^{PP}(\epsilon,V,T)=\frac{\mathcal{N}_l f_l+\mathcal{N}_r f_r+f_0(\Gamma_{\downarrow}\tau_{sf})^{-1}}{\mathcal{N}_l+\mathcal{N}_r+(\Gamma_{\downarrow}\tau_{sf})^{-1}}\\
f_{\uparrow}^{AP}(\epsilon,V,T)=\frac{e^{2d\delta\kappa}\mathcal{N}_l f_l+\mathcal{N}_r f_r+f_0(\Gamma_{\downarrow}\tau_{sf})^{-1}}{e^{2d\delta\kappa}\mathcal{N}_l+\mathcal{N}_r+(\Gamma_{\downarrow}\tau_{sf})^{-1}}\\
f_{\downarrow}^{AP}(\epsilon,V,T)=\frac{\mathcal{N}_l f_l+e^{2d\delta\kappa}\mathcal{N}_r f_r+f_0(\Gamma_{\downarrow}\tau_{sf})^{-1}}{\mathcal{N}_l+e^{2d\delta\kappa}\mathcal{N}_r+(\Gamma_{\downarrow}\tau_{sf})^{-1}},
\label{distributions}
\end{eqnarray}
where $\mathcal{N}_{l(r)}=\mathcal{N}(\epsilon\mp qV/2,T)$, $\mathcal{N}(\epsilon,T)=|\text{Re}[(\epsilon+i\gamma)/\sqrt{(\epsilon+i\gamma)^2-\Delta(T)^2}]|$ is the normalized  BCS density of states (DOS) in S, $\epsilon$ is the energy measured from the condensate chemical potential,
$\Delta(T)$ is the temperature-dependent superconducting energy gap, $\gamma$ is a smearing parameter \cite{gamma}, and $q$ is the electron charge. Furthermore, $f_{l(r)}=f_0(\epsilon \mp qV/2,T)$, $f_0(\epsilon,T)$ is the Fermi-Dirac distribution function, and $\tau_{sf}$ represents the characteristic spin relaxation time of quasiparticles in the N region. 
The spin-down tunneling injection rate is defined as $\Gamma_{\downarrow}=(\nu_N R_{\downarrow}\mathcal{A}t_Nq^2)^{-1}$, where $\nu_N$ is the DOS at the Fermi level in N, and $\mathcal{A}$ is the structure cross-sectional area \cite{gammaup}.
From Eqs. (1-4) it follows that when the injection rate greatly exceeds the relaxation rate, i.e., when $\Gamma_{\downarrow}\tau_{sf}\gg 1$, the distribution functions in the  N layer deviate considerably from $f_0(\epsilon)$.
In particular, in the absence of inelastic relaxation ($\Gamma_{\downarrow}\tau_{sf}\rightarrow \infty$), $f_{\uparrow}^{PP}(\epsilon,V)=f_{\downarrow}^{PP}(\epsilon,V)$, while in the opposite limit ($\Gamma_{\downarrow}\tau_{sf}\rightarrow0$) equilibrium is recovered. The electronic transport properties of the double spin valve are determined by the spin-dependent distribution functions. 
In particular, the quasiparticle current (e.g., evaluated at the left interface) in both configurations is given by $I^{PP(AP)}=\sum_{\sigma}I_{\sigma}^{PP(AP)}$ where
\begin{equation}
I_{\sigma}^{PP(AP)}(V)=\frac{1}{qR_{\sigma}}\int^{\infty}_{-\infty}d\epsilon \mathcal{N}_l(\epsilon)[f_l(\epsilon)-f_{\sigma}^{PP(AP)}(\epsilon)],
\label{current}
\end{equation}
while the differential conductance is given by $G^{PP(AP)}(V)=dI^{PP(AP)}/dV$.
\begin{figure}[t!]
\includegraphics[width=8.5cm]{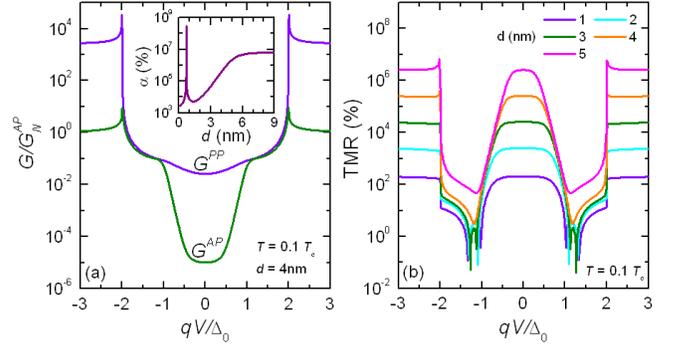}
\caption{\label{fig2} (Color) (a) Nonequilibrium differential conductance $G$ vs $V$ in the PP and AP configurations calculated for $d=4$ nm. The inset shows $\alpha$ vs $d$ (see text).
(b) Nonequilibrium TMR vs $V$ calculated for several barrier widths $d$.
In (a) and (b) we set $\Phi_0=0.8$ eV, $\delta E_{exc}=0.2$ eV, and $T=0.1T_c$.
}
\end{figure}

Figure 2(a) shows the differential conductance vs bias voltage $V$ in the PP and AP configurations calculated in the full nonequilibrium limit ($\Gamma_{\downarrow}\tau_{sf}\rightarrow \infty$) for $d=4$ nm at $T=0.1T_c$, where $T_c=(1.764 k_B)^{-1}\Delta_0$ is the superconducting critical temperature, $\Delta_0$ is the zero-temperature energy gap, and $k_B$ is the Boltzmann constant. 
In the following we suppose $T_c\ll T_{Curie}$, so to neglect any temperature dependence of $\delta E_{exc}$.
The curves are normalized to the  N-state nonequilibrium conductance in the AP configuration, $G_N^{AP}=2e^{2d\delta\kappa}[R_{\downarrow}(1+e^{2d\delta\kappa})]^{-1}$, and we set $\Phi_0=0.8$ eV and $\delta E_{exc}=0.2$ eV as representative parameters for Eu-based FIs \cite{Moodera2}. 
In both configurations $G$ is almost independent of voltage bias for $|V|>2\Delta_0/q$ (which corresponds to that in the normal state), and the ratio $G^{PP}/G^{AP}$ obtains $\simeq e^{2d\delta\kappa}/4$ for $e^{2d\delta\kappa}\gg 1$.
This large difference between $G^{PP}$ and $G^{AP}$ stems from the different barrier heights experienced by both spin species in the PP and AP configurations [see Fig. 1(c)] \cite{Worledge}.   
By contrast, $G$ is strongly bias-dependent for $|V|<2\Delta_0/q$ due to the presence of S electrodes.
This leads to a tunneling magnetoresistance ratio (TMR), defined as $\text{TMR}=(G^{PP}/G^{AP})-1$, which turns out to be largely tunable by changing the voltage bias across the structure.
The nonequilibrium TMR vs $V$ is displayed in Fig. 2(b) for $T=0.1T_c$ and for several values of $d$ \cite{thickness}.
We note first of all that TMR  is strongly enhanced as the barriers are made thicker, as expected for FI layers \cite{Worledge}.
As $G^{PP(AP)}(0)\simeq (\gamma/\Delta_0)G_N^{PP(AP)}$ at low temperature \cite{pekola}, $\text{TMR}(0)\simeq \text{TMR}_N$ where $\text{TMR}_N(\%)\simeq 25e^{2d\delta\kappa}$ is the N-state TMR.
Moreover, while TMR is nearly voltage-independent for $|V|>2\Delta_0/q$ approaching $\text{TMR}_N$, 
for each $d$ it can be tuned and suppressed by several orders of magnitude for $|V|<2\Delta_0/q$. 
In such a bias voltage range, by increasing barrier thickness yields to an enhancement of TMR suppression with respect to TMR$_N$.
For instance, with the given parameters,  TMR can be tuned already over \emph{four} orders of magnitude for $d\gtrsim 2$ nm, and stems from superconductivity-induced nonequilibrium in the N region.
 The TMR relative variation ($\alpha$), defined as $\alpha=[\text{TMR}(0)/\text{TMR}(\Delta_0/q)]-1$, is calculated in the inset of Fig. 2(a) vs $d$, and clearly shows this effect.
In addition, thicker barriers widen the voltage interval of TMR tunability, so that larger $d$ values are to be chosen in order to extend the TMR relative modulation with $V$.
Note that the finite-bias features appearing in the TMR characteristics of Fig. 2(b) reflect how the DOS of both S leads contribute to the total conductances at finite $V$ through Eqs. (1-4).
We stress that nonequilibrium as well as S electrodes are \emph{essential} elements for the observation of this spin valve effect. At equilibrium the distribution functions in N would be thermal and spin independent, while in the absence of superconductivity the nonequilibrium TMR would be virtually \emph{independent} of $V$ (at least for bias voltages of the order of $\Delta_0/q$).

The role of temperature is shown in Fig. 3(a) which displays the nonequilibrium TMR vs $V$ calculated for several values of $T$ and $d=4$ nm.
TMR$_N$ (dash-dotted line) is also shown for a comparison.
In particular, at the lowest temperature ($T=0.01T_c$), TMR turns out to be tunable with $V$ over about \emph{seven} orders of magnitude.
Then, by increasing the temperature leads to a reduction of this effect and, for higher $T$, TMR approaches TMR$_N$ which is independent of voltage.
The spin valve performance is thus enhanced at the lowest temperatures, although even  at $T=0.8 T_c$ the TMR can still be tuned over about two orders of magnitude.

\begin{figure}
\includegraphics[width=8.5cm]{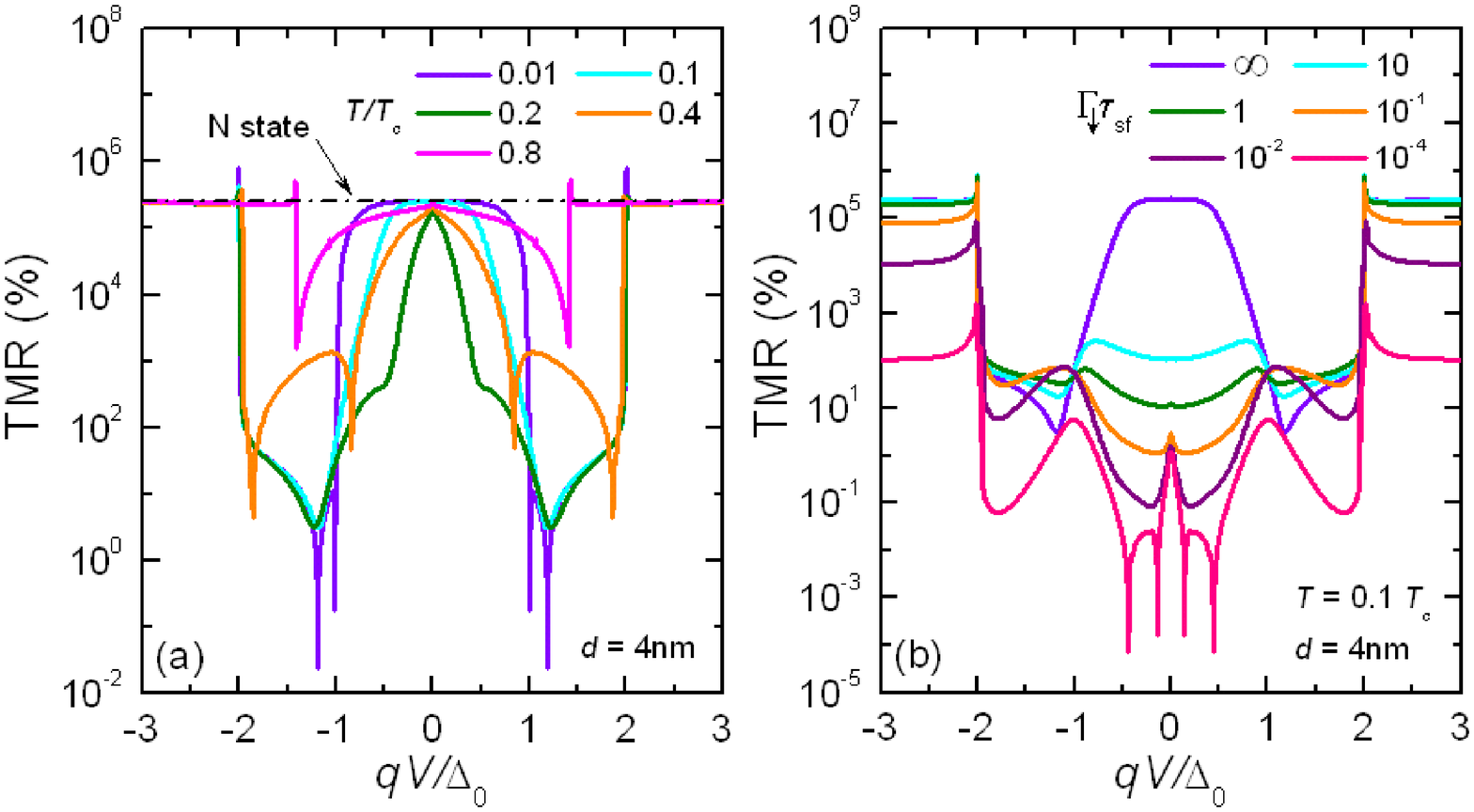}
\caption{\label{fig3} (Color) (a) Nonequilibrium TMR vs V calculated for several values of $T$. Dash-dotted line is TMR$_N$.
(b) TMR vs V calculated for different values of $\Gamma_{\downarrow}\tau_{sf}$ at $T=0.1T_c$. In (a) and (b) we set $\Phi_0=0.8$ eV, $\delta E_{exc}=0.2$ eV, and $d=4$ nm.
}
\label{fig3}
\end{figure}

The impact of inelastic relaxation in the N-region is displayed in Fig.~\ref{fig3}(b) which shows TMR vs $V$ calculated for different $\Gamma_{\downarrow}\tau_{sf}$ values at $T=0.1T_c$.
The general effect of decreasing $\Gamma_{\downarrow}\tau_{sf}$ is to suppress the TMR due to spin mixing which destroys the spin imbalance established in the N layer.
Notably, even in the presence of sizeable relaxation, e.g.,  $\Gamma_{\downarrow}\tau_{sf}=10^{-2}$, TMR values as large as $10^4\%$ and with a tunability over five orders of magnitude can, in principle, be achieved.

The spin-filtering nature of this setup can be quantified by the current polarization, defined as $\mathcal{P}_I=(I_{\uparrow}^{PP}-I_{\downarrow}^{PP})/(I_{\uparrow}^{PP}+I_{\downarrow}^{PP})=(e^{2d\delta \kappa}-1)/(e^{2d\delta \kappa}+1)$. This expression holds in both full nonequilibrium and equilibrium also in the absence of superconductivity, and is \emph{independent} of voltage. $\mathcal{P}_I\simeq 100\%$ can thus be obtained with a suitable choice of FIs and barrier widths \cite{Moodera3,LeClair,Moodera4,Santos}.

In light of applications, the device could be used for the implementation of storage cell elements or logic gates, where TMR can be  tuned continuously through an applied bias voltage. 
Magnetic field sensors or magnetic-field-controlled current switches  could be envisioned as well. 
In addition,  negligible power dissipation is intrinsic to the structure due to the presence of S electrodes, and makes this spin valve attractive for low-temperature applications.
With respect to practical realization, superconducting aluminum (Al) combined with Eu-based FIs appear as promising candidates \cite{Miao,Moodera2}, while  a conductor with a sufficiently long $\tau_{sf}$ is required for the N region. 
To this end heavily-doped Si, thanks to its  extremely long spin lifetime \cite{Castner,Jonker}, seems suitable and could be exploited, for instance, in a planar-like structure.  We recall that $\tau_{sf}$ as large as $\sim 1\,\mu$s has been measured in $n$-doped bulk Si (with $n\simeq 1\times 10^{19}$ cm$^{-3}$)  at 4K \cite{Castner}.
We derive here a general condition to retain a substantial TMR in the double spin valve. 
By referring to Fig. 3(c), a large tunable TMR is achievable provided $\Gamma_{\downarrow}\tau_{sf}\gtrsim10^{-4}$, i.e,  $R_{\downarrow}\mathcal{A}\lesssim10^4 \tau_{sf}/(\nu_F t_N q^2)$.
For the above cited case of Si with $\nu_N^{\text{Si}}\sim 6\times10^{45}$ J$^{-1}$m$^{-3}$ and $\tau_{sf}^{\text{Si}}=10^{-6}$ s \cite{Castner} we get 
$R_{\downarrow}\mathcal{A}\lesssim (65/t_N)\times 10^9$ $\Omega \mu$m$^2$
for the  spin-down specific resistance $R_{\downarrow}\mathcal{A}$ and $t_N$(nm), 
a condition that can be met  with a proper choice of material parameters \cite{Moodera2}.

In summary, we have proposed an out-of-equilibrium superconducting double spin valve based on ferromagnetic insulators. The structure is intrinsically simple and can be implemented with present-day technology. 
Besides providing very large and \emph{tunable} TMR values ($\sim 10^2\ldots 10^6\%$) which make it attractive for low-temperature magnetodevices, this setup may provide physical insight into the spin dynamics at the nanoscale, e.g., into the spin lifetime of nonmagnetic conductors.

We acknowledge F. Taddei for fruitful discussions and for critically reading the manuscript, and partial financial support from the NanoSciERA "NanoFridge" project.


\begin{thebibliography}{99}
\bibitem{interlayer}
	The presence of the N interlayer has the advantage to prevent direct exchange coupling between the two FIs.
\bibitem{gammaup}
	We have expressed Eqs. (1-4) as a function of $\Gamma_{\downarrow}$ which is the smaller injection rate, as $\Gamma_{\uparrow}=(\nu_N R_{\uparrow}\mathcal{A}t_Nq^2)^{-1}=e^{2d\delta\kappa}\Gamma_{\downarrow}$.
\bibitem{thickness}
	For simplicity we assume $\Phi_{\sigma}$ to be independent of $d$.
\end{thebibliography}
\end{document}